\documentclass[12pt]{article}
\usepackage{graphicx}

\begin{document}

\title{On Brownian Motion of Helium Ions in the
Ballistic Regime}

\author{A.Kleymenicheva, V.Shikin\\
Institute of Solid State Physics of RAS, 142432, Chernogolovka,
Russia}

\date{}
\maketitle

\begin{abstract}
Discussed in the paper is the possibility of introducing the
concept of Brownian motion of various mesoparticles in the
ballistic regime. The case in point is the effect of collisions
between thermal excitations in the liquid and the test
mesoparticle (allowing to trace its position) on the thermal
motion of the latter. The standard criterion for the brownian
nature of the particle motion assumes that the inequality
$l_s<R_i$ is satisfied and, consequently, a large number of
collisions with the averaged response of the mesoparticle to such
a bombardment (here $l_s$ is the mean free path of helium
excitations, $R_i$ is the effective mesoparticle radius). However,
the opposite limit $l_s>R_i$, which we refer to as the ballistic
regime, is also possible. It is characterized with the specific
features in the behavior of mesoparticles. The emphasis is made on
the already available evidences indicating this behavior in the
thermal motion of helium ions.
\end{abstract}
\vskip 3mm

The term Brownian is used to denote the thermal motion of various
mesoscopic particles (observed with appropriate techniques) due to
their interaction with thermal excitations in the environment. The
pioneering works [1, 2] on the diffusion-type behavior of brownian
particles laid the foundation of classical non-equilibrium
thermodynamics [3, 4].

A rather similar problem on the mobility $\mu_i$ of charged
particles in the external field contains two limiting cases. In
the first one, which is hydrodynamic in nature, when $l_s<R_i$,
the friction force is given the well-known Stokes formula $F_{st}$
[5] (here $l_s$ is the typical free path of the quasiparticles
which govern the medium viscosity, and $R_i$ is the radius of the
moving sphere)
$$
F_{st}= 6\pi R_i \eta V ,\eqno(1)
$$
where $V$ is the sphere velocity relative to the liquid whose
viscosity is $\eta$. The numerical coefficient in Eq. (1) may vary
according to the imposed boundary conditions (e.g., for the
Ryabchinsky problem this coefficient is 4 [5]).

The mobility $\mu_i$ of the ion of radius $R_i$ arising from the
condition $F_{st}=eE$, is
$$
\mu_i= V/E= e/(6\pi R_i \eta )  \eqno(2)
$$
where $E$ is the driving electric field pulling the charged sphere
through the viscous liquid.

In the opposite limit $l_s>R_i$, the so-called ballistic regime is
realized. Here the ion mobility is governed by the thermal
excitations kinetics. For ions in liquid helium in the roton
temperature range
$$
\frac{e}{\mu_{rot}}\simeq\sqrt{\frac{\pi}{2}}R_i^2v_{rot}.\eqno(3)
$$
Here $v_{rot}$ is the roton thermal velocity. The difference
between (2) and (3) is obvious. It is easily verified in the
experiments with helium ions (e.g., see Ref. [6]).

The question is what happens with thermal motion of mesoparticles
in the transition domain $l_s<R_i ~\to~l_s>R_i$? For helium ions
in liquid helium this crossover is easily realized already in the
temperature range of  $\le 2 K$ (just as for the the transition
from (2) to (3) in the problem of finding the particle mobility).

Qualitatively, the important ballistic alternative to the
diffusion-type brownian motion occurs in the dissipationless limit
$l_s~\gg~ R_i$, or, which is the same, $l_s \to \infty$. In that
case the ions or neutral impurity particles form a low-density gas
(suspension) in the solvent and obey Maxwell velocity distribution
since they practically behave as free particles. We believe that
the observation of free thermal motion of impurity particles of
any origin is sufficient to demonstrate the existence of the
ballistic regime in the brownian motion problem.

1. The simplest observable manifestation of the ballistic nature
of the thermal motion of impurities is the finite value of the
osmotic pressure $\delta P(c)\ne 0$ in various classic liquid
solutions,
$$
\delta P(c)= TC .\eqno(4)
$$
Here $T$ is the temperature and $C$ the impurity concentration.
Formally, the definition (4) arises when manipulating with ideal
chemical potentials of the solvent and the solved substance at the
semipermeable membrane [7]. Actually, the osmotic pressure is a
common practice when dealing with various (not necessarily ideal)
solutions.

2. More prominently the ballistic motion of ions manifests itself
in Doppler shift of optical spectra of the atoms of different
alkali metals artificially injected in liquid helium. The main
goal of these experiments is the study of the compressing effect
of surrounding helium on the external shells and, consequently, on
the luminescence spectrum of alkali atoms excited in the
superfluid liquid [8-10]. The expected shift of optical lines
should be accompanied with their broadening of different origin,
including thermal Doppler effect.

The frequency $\omega_0$ of radiation emitted by the atom whose
velocity component along the observation direction equals $v$ is
shifted, according to the Doppler principle, by $\omega_0 v/c$
$$
\omega=\omega_0(1+ \frac{v}{c}), \quad \mbox{or} \quad
v=(\omega-\omega_0)/c \eqno(4)
$$
where $c$ is the light velocity. Assume now that velocity
distribution of emitting atoms is defined by function $W(v)$.
Bearing in mind Eq. (4), one has
$$
I(\omega)d\omega= W(c,\frac{\omega-\omega_0}{\omega_0})
\frac{c}{\omega_0} d\omega \eqno(5)
$$

Foe the Maxwell distribution
$$
W(v)dv =\frac{1}{\sqrt{\pi}}\exp{[-(v/v_T)^2]}dv/v_T, \quad
v_T=\sqrt{2T/M_i}\eqno(6)
$$
Eq. (5) yields
$$
I(\omega)=\frac{1}{\sqrt{\pi}}\exp{\left[-\left(\frac{\omega-\omega_0}{\Delta
\omega_D}\right)^2\right]}\frac{d\omega}{\Delta \omega_D}, \quad
\Delta \omega_D=\omega_0 v_T/c \eqno(7)
$$
The frequency dependence $I(\omega)$ (7) is symmetric about
$\omega_0 $. The Doppler broadening width is defined by the
parameter $\Delta \omega_D$ which explicitly depends on the
emitter effective mass $M_i$.

The broadening of type (7) is known to yield one of the most
substantial contribution to the line width in optical spectra for
normal gaseous media [11]. For the low-temperature gas of alkali
atoms in liquid helium [10] the optical lines broadening
mechanisms are currently under study. Doppler broadening should
also be present among them.

This work was partly supported by the RFBR grant N 12-02-00229 and
the Program of the Presidium RAS ``Disordered systems''.

\end{document}